\documentclass[preprints,article,accept,moreauthors,pdftex]{Definitions/mdpi}
\usepackage{changes}

\firstpage{1}
\pubvolume{6}
\issuenum{4}
\articlenumber{46}
\pubyear{2022}
\copyrightyear{2022}
\externaleditor{{Academic Editors: Fabrizio Salvatore, Alessandro Cerri, Antonella De Santo and Iacopo Vivarelli} 
}
\datereceived{{1 August 2022} 
}
\dateaccepted{{13 September 2022}}
\datepublished{21 September 2022}
\hreflink{https://doi.\\org/10.3390/instruments6040046} 

\Title{Using Artificial Intelligence in the
Reconstruction of Signals from the PADME
Electromagnetic Calorimeter}

\TitleCitation{Using Artificial Intelligence in the
Reconstruction of Signals from the PADME
Electromagnetic Calorimeter}

\Author{Kalina Dimitrova *\orcidA{} and on behalf of the PADME collaboration $^{\dagger}$ }

\AuthorNames{Kalina Dimitrova}

\AuthorCitation{Dimitrova, K.; 
on behalf of the PADME Collaboration.}

\address[1]{
Faculty of Physics, Sofia University ``St. Kliment Ohridski'', 5 J. Bourchier Blvd.
, 1164 Sofia, Bulgaria}

\corres{\hangafter=1 \hangindent=1.05em \hspace{-0.82em} Correspondence: kalina@phys.uni-sofia.bg}

\firstnote{\hangafter=1 \hangindent=1.05em \hspace{-0.82em} The PADME collaboration: A.P. Caricato, M. Martino, I. Oceano, F. Oliva, S. Spagnolo (INFN Lecce and Salento~Univ.), G.~Chiodini (INFN Lecce), F. Bossi, R. De Sangro, C. Di Giulio, D. Domenici, G.~Finocchiaro, L.G.~Foggetta, M. Garattini, A. Ghigo, P. Gianotti, I. Sarra, T. Spadaro, E. Spiriti, C. Taruggi, E.~Vilucchi (INFN~Laboratori Nazionali di Frascati), V. Kozhuharov (Faculty of Physics, Sofia Univ. ``St.~Kl.~Ohridski'' and INFN~Laboratori Nazionali di Frascati), S. Ivanov, Sv. Ivanov, R. Simeonov (Faculty of Physics, Sofia Univ. ``St.~Kl.~Ohridski''), G. Georgiev (Sofia Univ. ``St.~Kl.~Ohridski'' and INRNE Bulgarian Academy of Science), F.~Ferrarotto, E.~Leonardi, P. Valente, A. Variola (INFN Roma1), E. Long, G.C. Organtini, G. Piperno, M. Raggi (INFN Roma1 and ``Sapienza'' Univ. Roma), S. Fiore (ENEA Frascati and INFN Roma1), V. Capirossi, F.~Iazzi, F.~Pinna (Politecnico of Torino and INFN Torino), A. Frankenthal (Princeton University). 
}

\abstract{The PADME apparatus was built at the Frascati National Laboratory of INFN to search for a dark photon ($A’$) produced via the process $e^+ e^- \rightarrow A’\gamma$. The central component of the PADME detector is an electromagnetic calorimeter composed of 616 BGO crystals dedicated to the measurement of the energy and position of the final state photons. The high beam
particle multiplicity over a short bunch duration requires reliable identification and measurement of overlapping signals. A regression machine-learning-based algorithm has been developed to disentangle with high efficiency close-in-time events and precisely reconstruct the amplitude of the hits and the time
with sub-nanosecond resolution. The performance of the algorithm and the sequence of improvements leading to the achieved results are presented and discussed.}


\begin{document}



\section{Introduction}
{In recent years, the~search for an explanation of the Dark Matter phenomenon has led to the development of various hypotheses for an extension of the Standard Model, e.g.,~Weakly Interacting Massive Particles (WIMPs) \cite{bib:WIMPs}. However, the~non-observation of new states with mass
in the order of 100 GeV led scientists to explore other Dark Matter explanations.}
The main goal of PADME (Positron {Annihilation} into Dark Matter Experiment) \cite{bib:PADME-idea} is to search for the dark photon $A'$, a~hypothetical gauge boson connecting the dark and the visible sector. In~the case of non-vanishing interaction strength $\alpha '$ with the electrons, $A'$ can be produced in the annihilation process of beam positrons with electrons from the target:
\begin{linenomath}
\begin{equation}
e^{+}e^{-}\rightarrow A' \gamma.
\end{equation}
\end{linenomath}

Knowing 
the four-momenta
of the beam's positrons, the~electrons at rest and the photon produced in the process, the~missing mass of the dark photon can be calculated:
\begin{linenomath}
\begin{equation}
M^2_{miss} = (P_{e^+}+P_{e^-}-P_{\gamma})^2.
\end{equation}
\end{linenomath}

The positron beam provided by the DA$\Phi$NE LINAC~\cite{bib:dafne} can reach energies up to 550~MeV, providing a limit for the missing mass of 23.7 MeV, and~is composed of bunches with a 50 Hz rate. Each bunch contains about $2\times10^4$ particles and its length can be varied with typical values of 200--300 ns.

The two main processes contributing to the background, are the annihilation $e^{+}e^{-}\rightarrow \gamma \gamma (\gamma)$ and Bremsstrahlung events $e^+N\rightarrow e^+N\gamma$.

To suppress the background from $e^{+}e^{-}\rightarrow \gamma \gamma (\gamma)$,
the PADME experiment should have high photon
detection efficiency, while for the rejection of
$e^+N\rightarrow e^+N\gamma$, the~radiating positron
should be detected and a reliable matching in time between the positron
and the emitted photon should be~assured.

\textls[-5]{The initial studies based on a full Geant4~\cite{GEANT4:2002zbu} simulation
indicate that the PADME experiment can reach sensitivies
in $\alpha '$ down to $10^{-8}$ \cite{Raggi:2015gza} with high-efficiency detectors (greater than 99\%)
and time resolution better than 1 ns.
In addition, due to the high bunch
multiplicity, double-pulse separation capabilities are required for each of the chosen~detectors.}


\section{The Padme~Experiment}

\begin{figure}[H]
\includegraphics[width=.98\columnwidth]{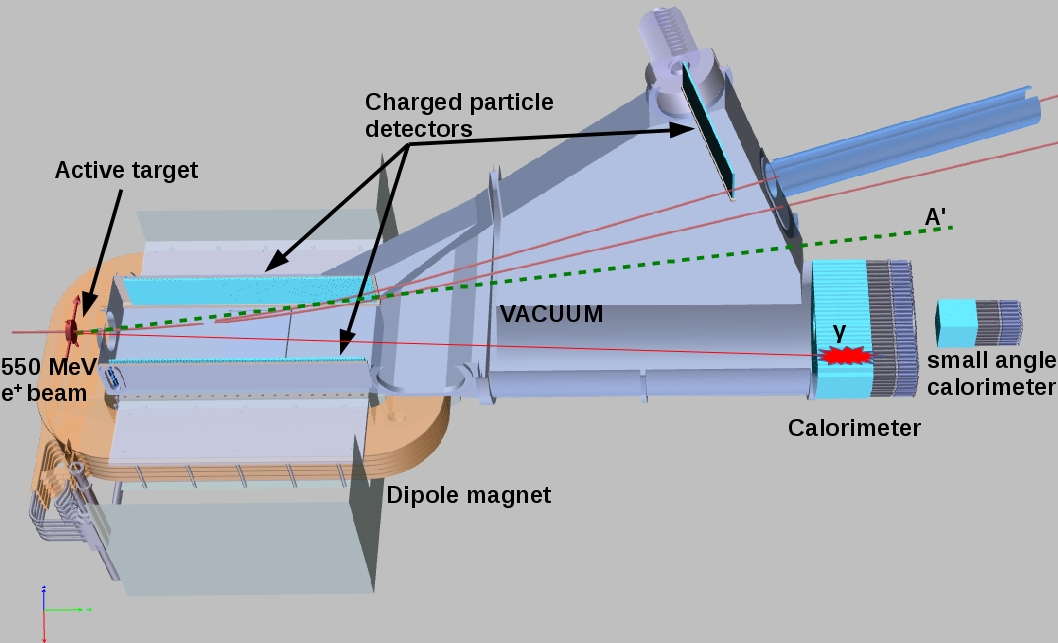}
\caption{Outline 
of the PADME experiment\added{.}\label{fig1}}
\end{figure}

A sketch of the PADME experiment is shown in {Figure} \ref{fig1}.
A short description of its major detector components~\cite{bib:padme-tdr} follows.
\vspace{-6pt}

\subsection{Active~Target}
The target~\cite{bib:padme-target}
is composed of polycrystalline diamond (Z = 6) since low Z is required
to increase the
annihilation to bremsstrahlung cross-section ratio. The~target has a {100~$\upmu$m} thickness and 20 mm width and length. Apart from providing the target for the annihilation process, it also measures the beam's multiplicity and XY profile. For~this reason 16~horizontal and 16~vertical graphite electrodes of 1 mm width are engraved onto the target using an excimer~laser.



\subsection{Charged Particle~Detectors}
Three sets of detectors
register the charged particles. {The beam positrons may lose energy in
the target and produce Bremsstrahlung photons, detected by the electromagnetic calorimeter, which need to be rejected. This is achieved by coinciding these photons with the particles that produced them. These particles are detected by the positron and high energy positron vetoes. 
In case the $A'$ decays
into an $e^+e^-$ pair, the~electron will be registered by the electron veto.} All {three charged particle detectors} are composed of {$10\times10\times178$ mm$^3$} plastic scintillators with WLS fibers coupled to $3 \times 3$ mm$^2$ Hamamatsu S13360 silicon photomultipliers with 25 $\upmu$m pixel size and are placed in $10^{-5}$ mbar vacuum. The~positron and the electron vetoes are located inside the magnet and are composed of 90 and 96 scintillating bars, respectively. Both detect particles with momenta between \mbox{50 and 450 MeV.}
The high energy positron veto is located next to the beam exit window and is composed of 16 scintillating bars, with~scintillation light read out on both sides. It allows the detection of positrons with momenta between 450 and 500~MeV.

The charged particle detectors segmentation {provides} measurement of the $e^+/e^-$ momentum with a resolution of $\approx$5\%. The~time resolution is 700 ps~\cite{bib:padme-veto}.


\subsection{Calorimeters}
\begin{figure}[H]
\includegraphics[width=0.6\columnwidth]{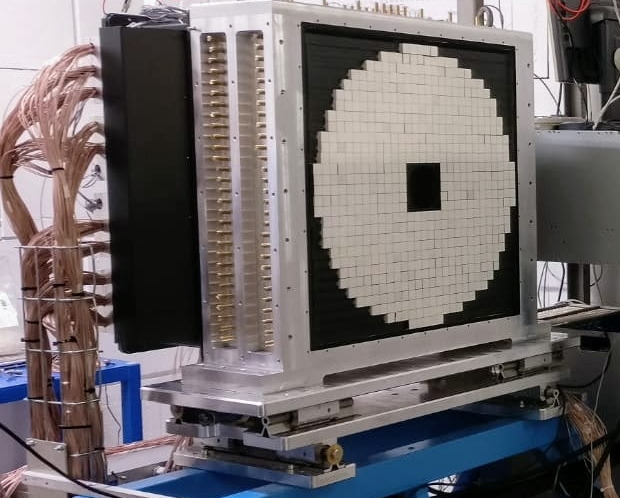}
\caption{The PADME Electromagnetic~Calorimeter.\label{fig4}}
\end{figure}

The PADME calorimetric system is composed of an Electromagnetic Calorimeter (ECal) and a small-angle calorimeter (SAC).
The ECal (Figure \ref{fig4}) is composed of 616~BGO cystals measuring  $2.1 \times 2.1 \times 23$ cm$^3$, connected to HZC 1912 photomultipliers. The~optical isolation of the crystals is achieved by covering
them with diffuse reflective TiO$_2$ paint and additionally with 50–100~$\upmu$m thin black
Tedlar foils. The~ECal is placed 3.45~m away from the target and has a radius of 29 cm, thus achieving an angular coverage between \mbox{15 and 84 mrad.} The~lower limit is due to a square hole in its center, which is covered by the SAC. The~scintillation light decay time is 300 ns. Calibration was performed both with a $^{22}$Na source before constructing the calorimeter and then subsequently with cosmic rays. The~energy resolution is $\sim$7\% at $E_{\gamma} =100$ MeV~\cite{bib:padme-ecal}.

The Small Angle Calorimeter (SAC) \cite{bib:padme-sac} is located downstream of the ECal
and mainly detects photons produced by Bremsstrahlung events.
To suppress this type of background,
the data from the SAC is
matched in time with the
data from the charged particle detectors.
In addition, the~SAC detects photons from multiphoton annihilation events.
The SAC is composed of 25 {PbF$_2$} crystals measuring $ 3 \times 3 \times 14$ cm$^3$ and covers an angle between 0 and 15~mrad.

\subsection{Readout~System}
The PADME Data Acquisition System
consists of 29 CAEN V1742 ADC boards, each equipped with 32 analog and 2 trigger input channels. The~V1742 switch capacitor digitizer employs the DRS4 chip, capable of sampling
the input signal at 750 MS/s, 1 GS/s, 2.5 GS/s, and~5 GS/s.
Complete waveforms of 1024 samples for each channel are recorded upon a beam-based trigger signal.
In the case of the electromagnetic calorimeter which is sampled at 1 GS/s,
this corresponds to a {$\approx$1~$\upmu$s} recorded~waveform.


\section{Application of Neural Networks for Waveform~Description}

The high multiplicity of the positron beam in combination with the short bunch
duration leads to many overlapping or close-in-time signals recorded in a single event.
One way to solve this problem is to use neural networks (NNs) to count the signals {from the electromagnetic calorimeter} in each event, to~identify and separate overlapping signals, and~to extract signal parameters---the arrival times of the individual signals and their amplitudes. 
To train the NNs, {an} event simulation was developed. Each event represents a waveform of 1024 samples, for~which the number of signals as well as the individual signal parameters can be varied.
Various sets of waveforms $A(t)$ were generated, containing a random number of signals with shapes defined by the subtraction of two exponents
\begin{linenomath}
\begin{equation}
A(t)=A_0(e^{-(t-t_0)/\tau_1}-e^{-(t-t_0)/\tau_2}), ~~~~t\geq t_0,
\end{equation}
\end{linenomath}
where $t_0$ is the arrival time of the signal, $\tau_1$ is the signal decay time, taken to be 300 ns{,} and $\tau_2$ is related to the signal rise time, taken to be 10 ns. {These values are typical for the PADME PMT + BGO crystal assembly.}
$A_0$ is the signal amplitude parameter, chosen to follow a Gaussian distribution. {The arrival time follows a uniform distribution with a minimum value of $t_0=100$ ns, to~account for the trigger specifications.}
For the training of the networks presented here, a~mean value $A_0 = 200$ mV with {$\sigma = 200$ mV} was used and an additional lower limit $A_0 > 20$ mV was set. {The signals from the ECal are digitised by an ADC with 1 V dynamic range, which should be sufficient for the maximal energy cell within an electromagnetic shower. Since the selected photon energy is between 50~MeV and 450 MeV, a~200 mV mean amplitude was chosen to increase the training statistics to signals corresponding to the 100 MeV range.} All {waveforms} include {a} Gaussian noise {with} a mean value of 10 mV {added in each bin}. A~predefined maximum number of {four} signals was used for all generated waveforms{. Many of the events, recorded by the ECal contain only one or two signals, however, there are events with more recorded signals which requires the inclusion of such cases
in the training.} {Different NNs were trained on the thus generated events and }each network is trained on 100,000 events.




For the implementation of all neural networks presented in this study and for the output analysis were used the ROOT~\cite{ROOT}, TensorFlow~\cite{bib:TF} and Keras~\cite{bib:Keras} frameworks.
Three different convolutional neural networks (CNNs) {\cite{bib:CNN}} were developed{, starting with the classification task of dividing the events into categories based on the number of signals in
them and moving to regression tasks with the aim of signal parameter estimation}.

\begin{figure}[H]
\includegraphics[width=0.70\columnwidth]{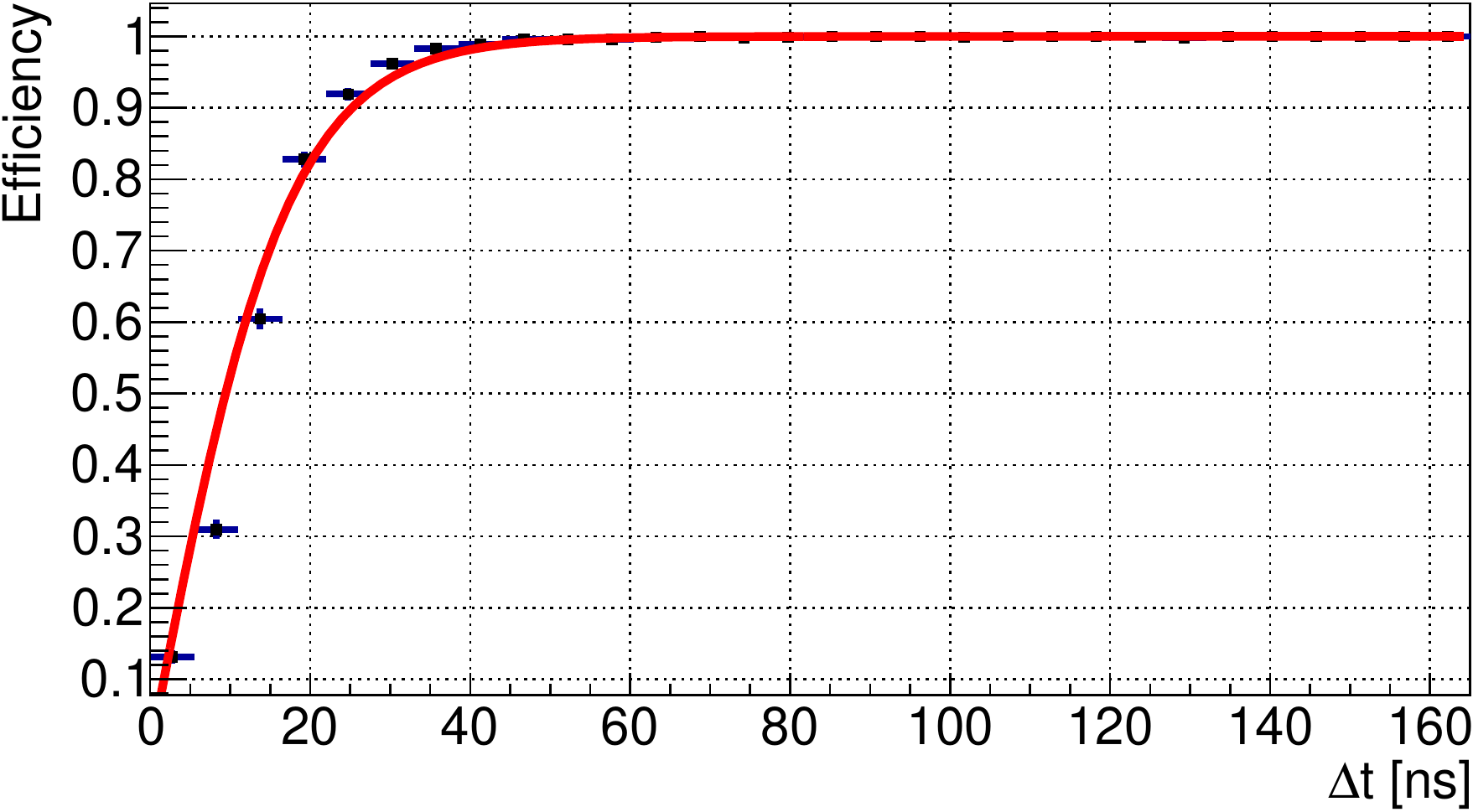}
\caption{Efficiency of a signal-counting CNN {as a function of} the time difference between two signals. {A sigmoid curve for the efficiency Eff 
$=2\cdot(\frac{1}{(1 + exp(-\Delta t/70)}  - 0.5)$ is fitted (red).} For signals with $\Delta t = 10$ ns, the~efficiency is 50\%. The~efficiency reaches 100\% for $\Delta t > 50$ ns.\label{fig7}}
\end{figure}

The first CNN performs a classification task aimed at counting the number of signals in each waveform. {It consists of a single convolutional layer, followed by three fully connected layers. This network} was trained using labels containing only information about the number of signals in each event. The~obtained model was then applied for the simulated set of events with two generated signals and the output was compared to the true labels. Figure~\ref{fig7} shows the efficiency of the signal counting as a function of $\Delta t = |t_2 - t_1|$ where $t_2$ and $t_1$ are the arrival times of the two signals. The~efficiency is 50\% for $\Delta t = 10$ ns and 100\% for $\Delta t > 50$ ns.

The estimation of signal parameters requires the development of networks with more complex architectures. Convolutional autoencoders {\cite{zhang2018better}} can be used for extracting useful data from~waveforms.

\begin{figure}[H]
\includegraphics[width=0.95\linewidth]{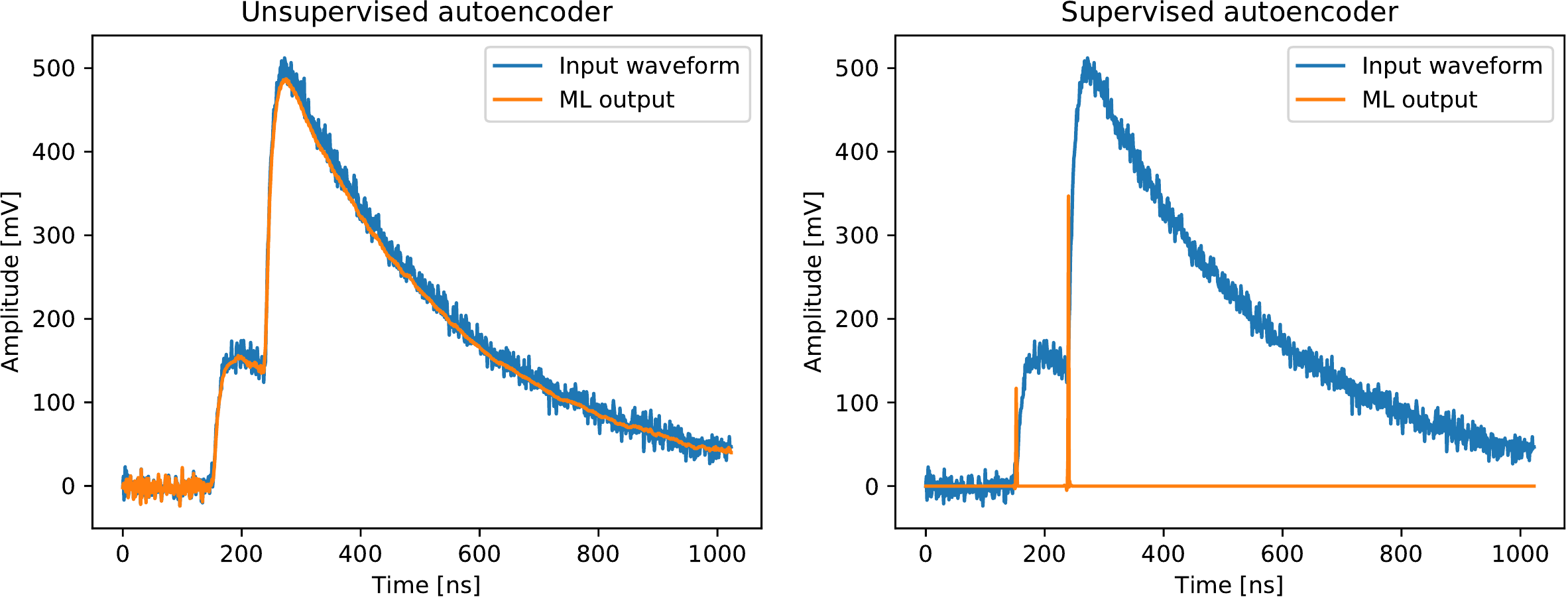}
\caption{Autoencoder 
NN output and the original {event with two signals}. (\textbf{Left})  {classical} autoencoder. (\textbf{Right}) {modified} autoencoder.\label{fig89}}
\end{figure}

An autoencoder was developed with the targeted output replicating the original waveform. The~neural network architecture consists of three convolutional layers followed by three deconvolutional layers with the same parameters and a single final deconvolutional layer for setting the output dimensions.
Figure~\ref{fig89} {(left)} shows an example of the original and the output waveform {for an event with two signals}.
It is observed that such networks can reproduce the signals, which represents the main motivation for applying this architecture for signal parameter estimation through supervised~learning.

A modified autoencoder trained on labeled data was developed with training based on output arrays of the same length as the input waveforms. On~the positions of signal arrival, the~value is set to the signal amplitude and all other values are set to 0. An~example of the output of the {modified} autoencoder and its corresponding waveform is presented in Figure~\ref{fig89} {(right)}.

To assess the modified autoencoder output and compare it with the original data labels, a~reconstruction {algorithm} was developed.
The data labels contain an array of 1024~numbers with a single non-zero amplitude value at the arrival time $t_0$ for each generated signal.
The NN output gives multiple amplitude values over a number of time positions for each recognised signal, usually with a maximum on {the most probable} position and decreasing values on both of its sides.
The reconstruction algorithm locates the maximum and adds the values of the three positions before it and the three positions after it to the maximum value.
The result is taken to be the amplitude of the reconstructed signal and the time position of the maximum is taken to be the arrival time of this signal.
The reconstruction is also applied to the original data labels and the results for the reconstructed output are then compared to them.
This allows both for comparison of the amplitude value of the original and reconstructed signal, as~well as evaluation of the arrival time and analysis of the neural network~efficiency.

\section{Signal Parameter~Reconstruction}

The probability for a signal to be discriminated
and the accuracy of the reconstructed signal parameters were studied.
The modified autoencoder model was trained on a set of 100,000 events with up to {four} signals each and was applied to a statistically independent {test }set, again with 100,000 events containing up to {four} signals.

\subsection{Time~Reconstruction}
\begin{figure}[H]
\includegraphics[width=.45\textwidth]{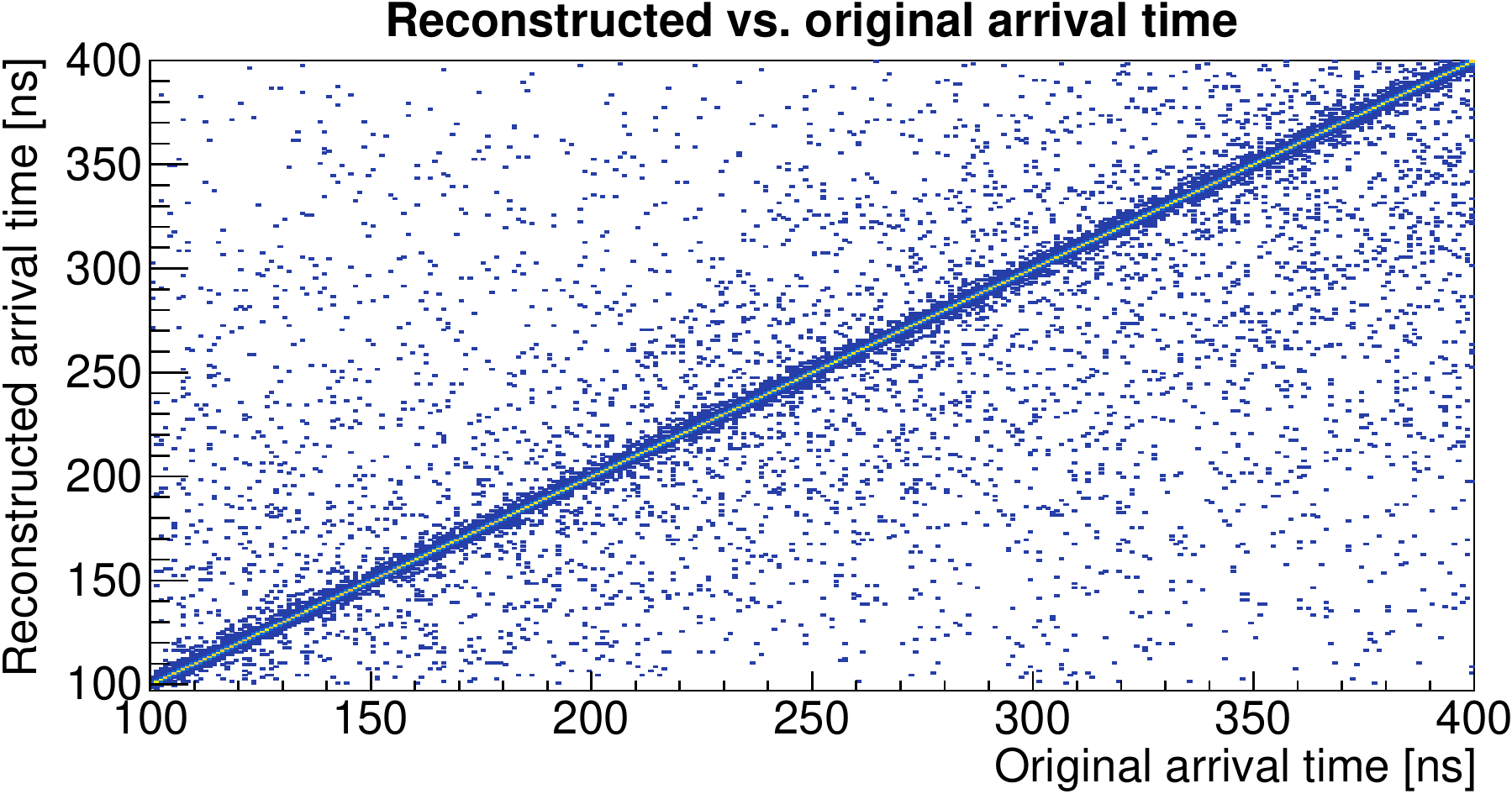}
\includegraphics[width=.45\textwidth]{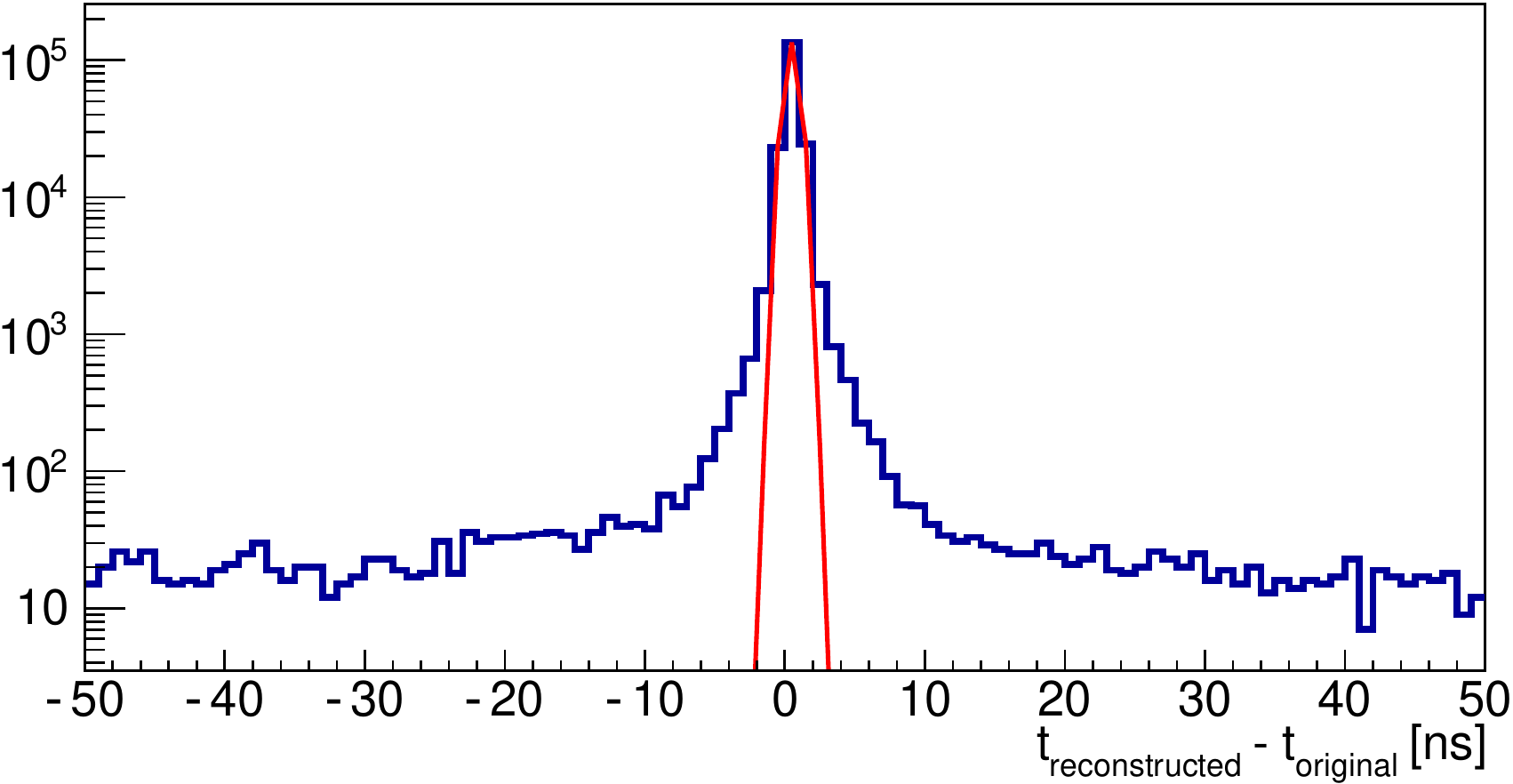}
\caption{\textls[-15]{(\textbf{Left}) 
{Reconstructed vs. original time of arrival for all events in the test set.} (\textbf{Right}) Distribution of the signals according to the difference between the reconstructed and the original signal {arrival time. The~red curve represents a Gaussian distribution with $\sigma \approx 520$ ps and mean value of 0}.}\label{fig1314}}
\end{figure}

To study the double-pulse separation abilities of the machine learning algorithm, each {simulated} signal is associated with the closest-in-time one from the NN output. 
The left panel in Figure~\ref{fig1314} represents the difference between the reconstructed and the original time of {the} signal arrival. The~distribution of this difference ({Figure}~\ref{fig1314}, right panel) is symmetric, with~$\sigma$ $\sim$ 520 ps and RMS $\sim$ 3.2 ns if assumed Gaussian, however, non-Gaussian tails do~exist.  

A signal is considered successfully identified if the difference between the original and reconstructed output is less than 2~ns.

\subsection{Signal~Recognition}
\begin{figure}[H]
\includegraphics[width=0.62\columnwidth]{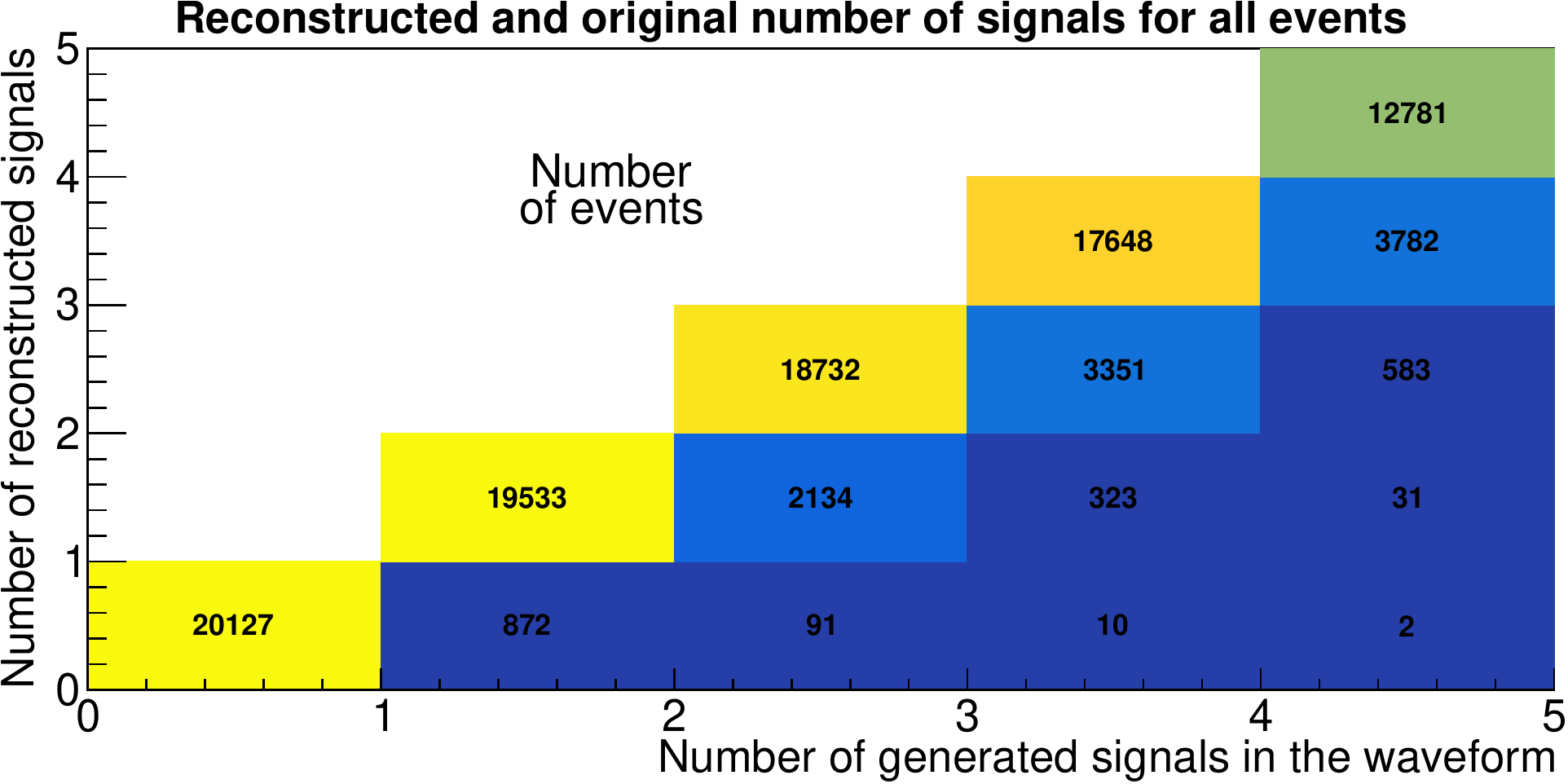}
\caption{Number 
of reconstructed signals versus the original number of generated signals in the test~sample.\label{fig10}}
\end{figure}

Figure~\ref{fig10} shows all of the events {in the test set}, divided into bins based on the number of reconstructed signals and the originally generated {ones}. Ideally, these two numbers should be the same for all events.
However, two major factors influence signal discrimination: a small difference in signal arrival time may cause two or more signals to be merged
into one and signals with small amplitudes may not be identified above the~noise.

\begin{figure}[H]
\setkeys{Gin}{width=0.45\linewidth}
\includegraphics{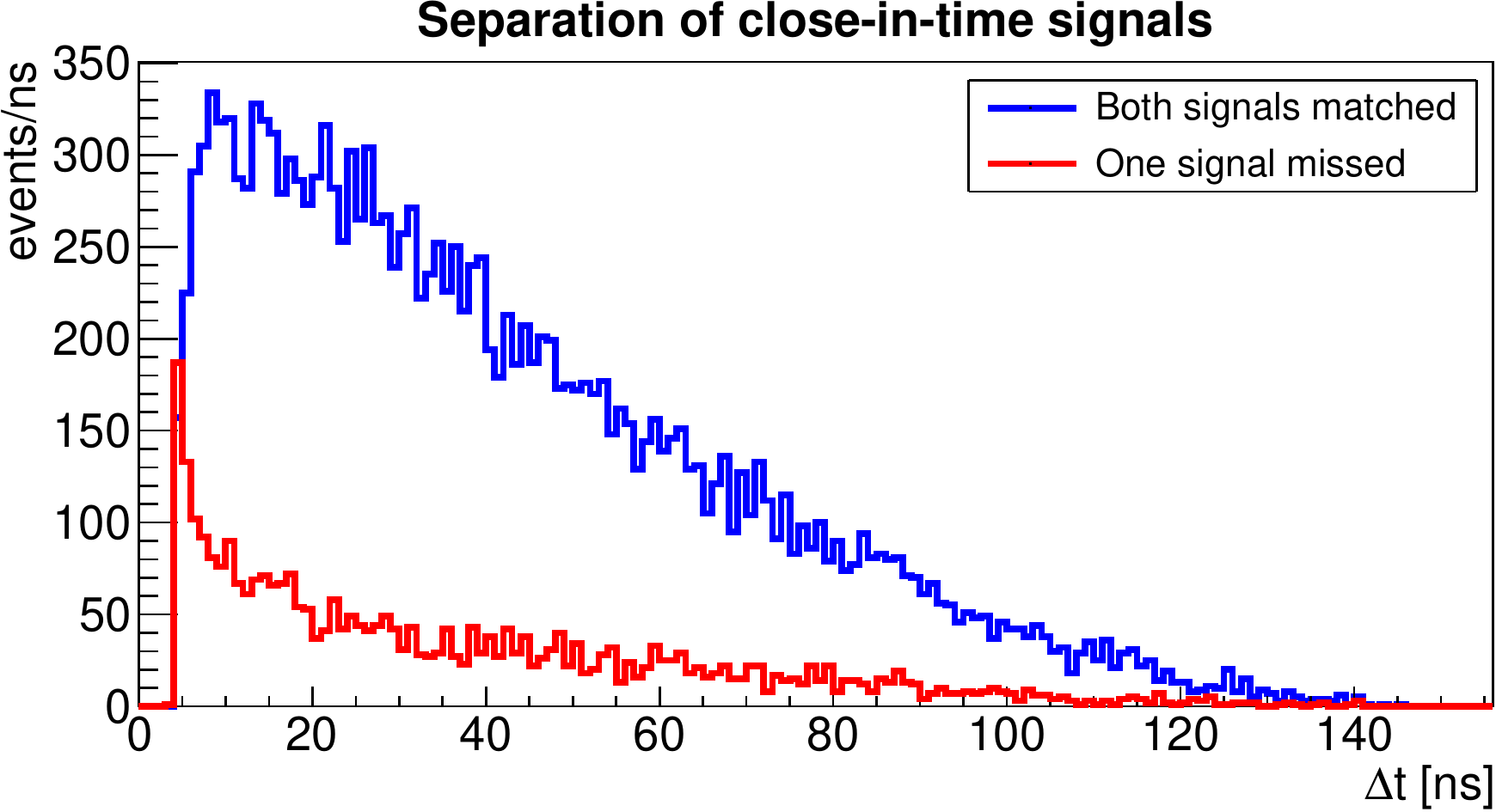}
\includegraphics{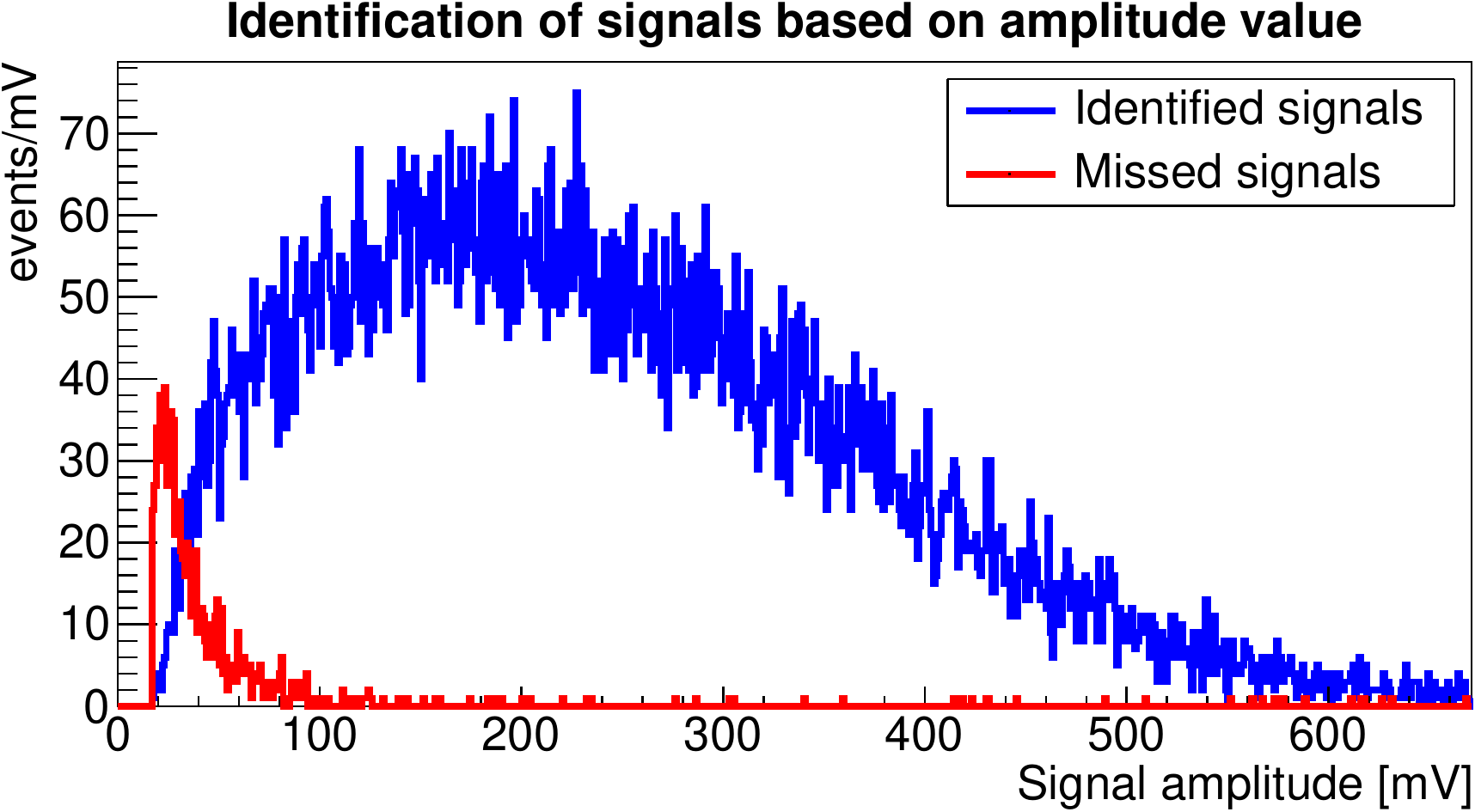}
\caption{(\textbf{Left}) Matched (blue) and missed (red) events as a function of the arrival time difference $\Delta t$ for events with two generated signals. (\textbf{Right}) Matched (blue) and missed (red) events as a function of the amplitude value for events with one generated signal. Missed events with high time differences are due to small~amplitudes.\label{fig1112}}
\end{figure}

Signals with time differences less than 10 ns are merged into a single hit
and most events with amplitudes smaller than 50 mV are not likely to be identified\added{,}
which results in decreased efficiency, as~seen in {Figure} \ref{fig1112}.

\unskip

\subsection{Amplitude~Reconstruction}
\begin{figure}[H]
\setkeys{Gin}{width=0.45\linewidth}
\includegraphics{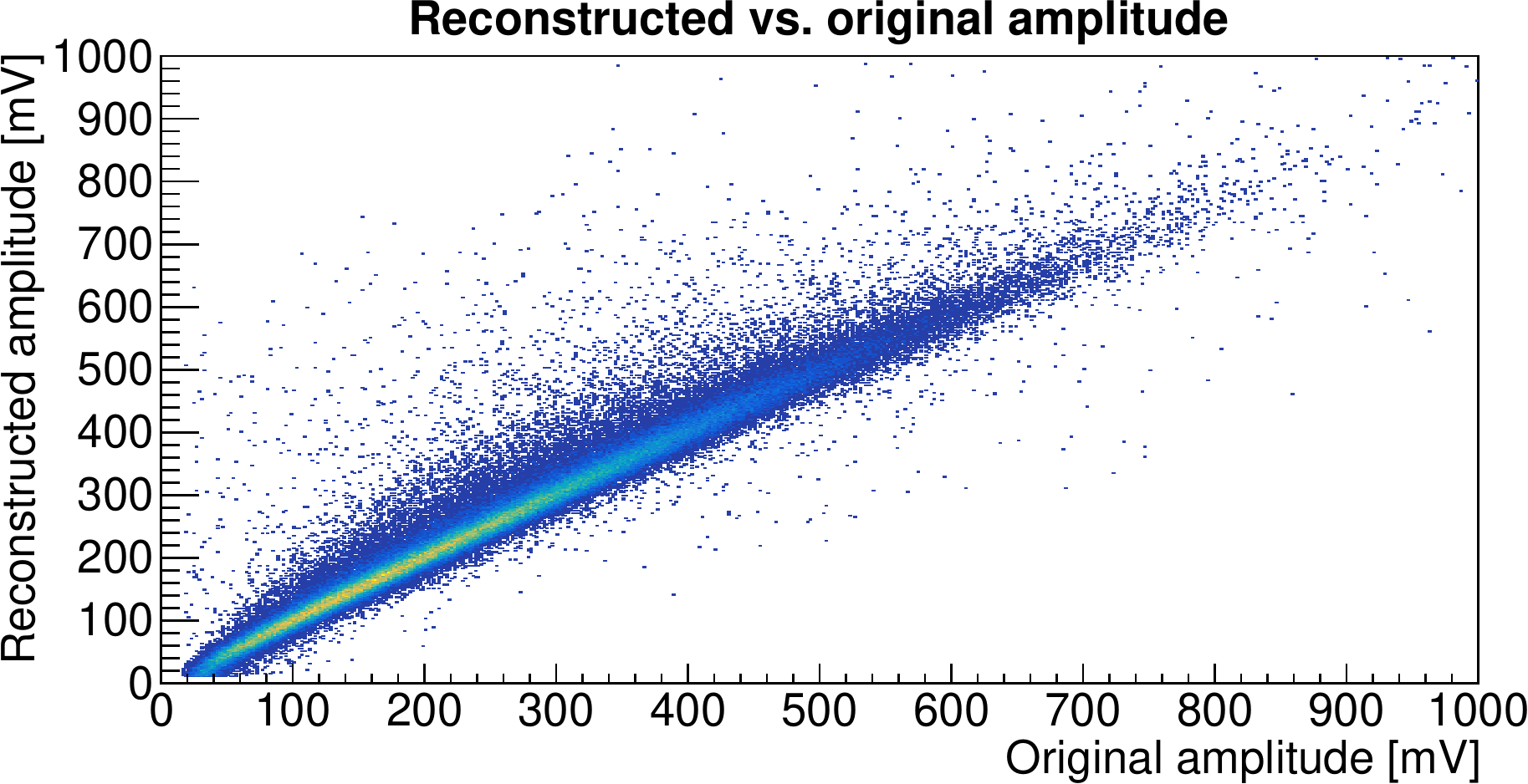}
\includegraphics{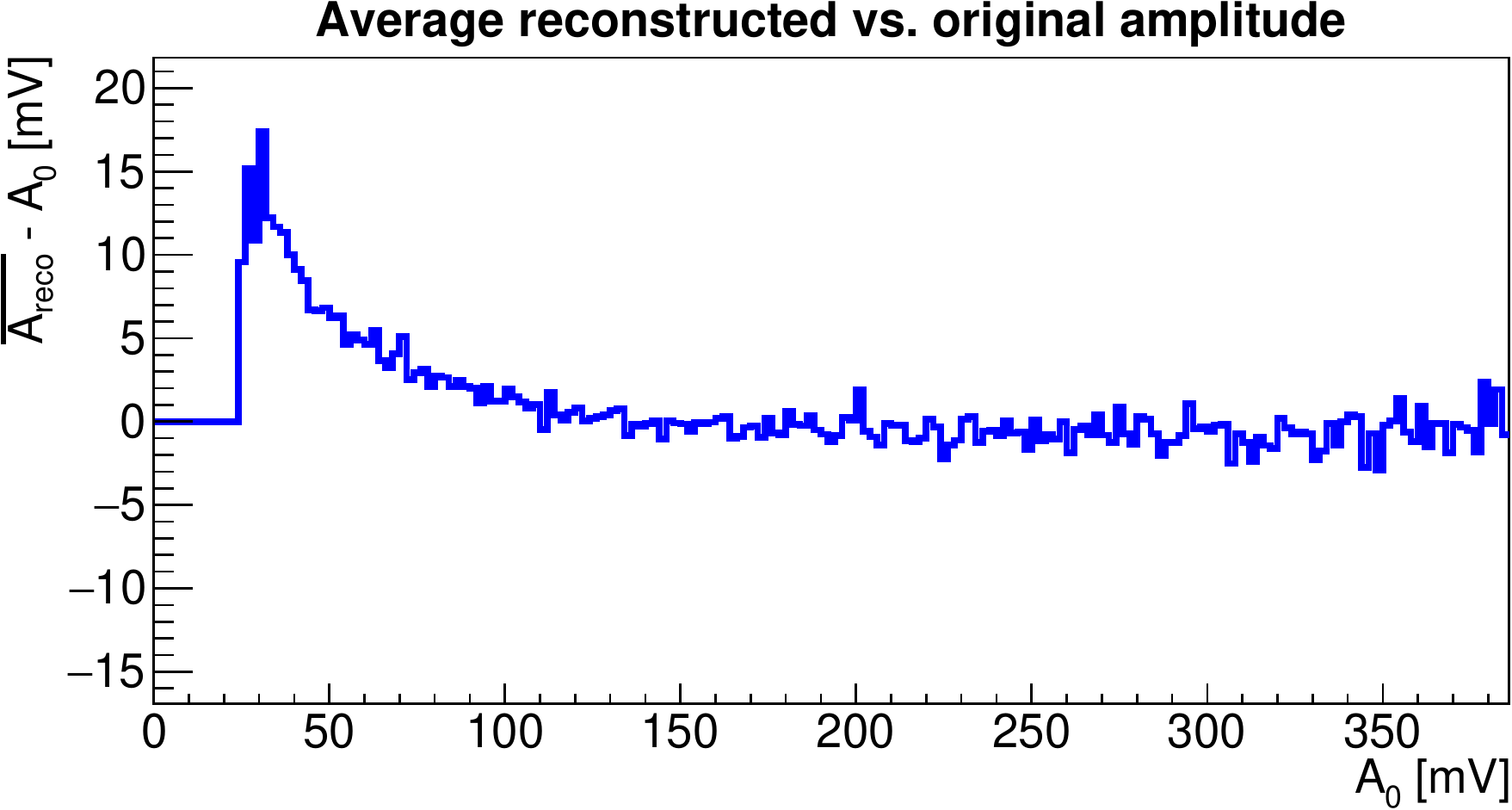}
\caption{(\textbf{Left}) All events based on their reconstructed amplitude value and the original one  (\textbf{Right}) Difference between the average value of the reconstructed amplitude $\overline{A_{reco}}$
and the original value $A_0$ for each original value $A_0${, divided into 2 mV bins}. It can be observed that for small amplitudes the reconstructed value is higher than the original~one. \label{fig1516}}
\end{figure}

The developed CNN provides reconstruction of the signal amplitude values.
The reconstructed versus the original amplitudes for the successfully identified signals are shown in the left panel of Figure~\ref{fig1516}.
The correlation is well pronounced. To~quantitatively compare the quality of signal identification, the~average difference between the reconstructed and the true amplitude value for each 2 mV interval of generated amplitude is shown in the right panel of Figure~\ref{fig1516}.
\vspace{-6pt}

The inaccuracy of the small amplitude values can be compensated for by a dedicated calibration of the NN output.

\section{Conclusions}
{The high particle rate in the PADME calorimeter requires implementation of advanced reconstruction algorithms to achieve less than 1 ns time resolution of the reconstructed showers.}
Machine learning methods were applied for the successful identification of calorimeter signals and for the extraction of signal parameters.
A CNN with a single convolutional layer was used
to count the number of signals in an event.
A modified CNN autoencoder was probed for the estimation of signal parameters---arrival time and amplitude. 
The performance of the networks was assessed through a specially designed algorithm, comparing the network output with the original data labels.
The CNN provides time reconstruction with $\sim$500 ps time resolution and the amplitude is reconstructed in the 30--700 mV range.
There is an inaccuracy for amplitudes less than 100 mV, which can be solved by an additional calibration of the NN output or by developing networks with modified architectures specifically targeting small amplitudes.

\vspace{6pt}

\authorcontributions{
The presented material is
a result of the joint work of the PADME collaboration.  All authors have read and agreed to the published version of the manuscript.
}

\funding{This work was partially supported by 
BNSF: KP-06-D002\_4/15.12.2020 within MUCCA, CHIST-ERA-19-XAI-009 and by LNF-SU~MoU.}

\dataavailability{Not applicable.
}



\conflictsofinterest{The authors declare no conflict of~interest.
}


\begin{adjustwidth}{-\extralength}{0cm}

\reftitle{References}



%


%
%
%
\end{adjustwidth}
\end{document}